# Physical Time and Human Time

George F. R. Ellis

Mathematics Department, University of Cape Town, South Africa.
email: george.ellis@uct.ac.za

The New Institute, Hamburg, Germany
email: george.ellis@thenew.institute

ORCID number 0000-0001-8484-0629

**ABSTRACT**. *This paper is a comment on both Bunamano and Rovelli [1] and Gruber et al [2] and which discuss the relation between physical time and human time. I claim here, contrary to many views discussed there, that there is no foundational conflict between the way physics views the passage of time and the way the mind/brain perceives it. The problem rather resides in a number of misconceptions leading either to the representation of spacetime as a timeless Block Universe, or at least that physically relevant universe models cannot have preferred spatial sections. The physical expanding universe can be claimed to be an Evolving Block Universe with a time-dependent future boundary, representing the dynamic nature of the way spacetime develops as matter curves spacetime and spacetime tells matter how to move. This context establishes a global direction of time that determines the various local arrows of time. Furthermore time passes when quantum wave function collapse takes place to an eigenstate; during this process, information is lost. The mind/brain acts as an imperfect clock, which coarse-grains the physical passage of time along a world line to determine the experienced passage of time, because neural processes take time to occur. This happens in a contextual way, so experienced time is not linearly related to physical time in general. Finally I point out that the Universe is never infinitely old: its future endpoint always lies infinitely faraway in the future.*

**Acknowledgements:** I thank the University Research Committee of the University of Cape Town for financial support.

**Disclosure of potential conflicts of interest:** The author has no conflicts of interests.

## 1. Introduction: Two incompatibilities

This paper is a response to both *"Bridging the neuroscience and physics of time"* by Buanomano and Rovelli ([1], hereafter BR), and *"Physical Time Within Human Time"* by Gruber, Block, and Montemayor ([2], hereafter GBM). Both discuss the relation between the nature of time as viewed by physics and as viewed from considerations of brain function. The former rejects both static eternalism (the block universe view) and global presentism, and rather proposes a multilayered concept of time for both physics and the mind. The latter quotes various claims that the perceived passage of time (the *Flow of time*, or FOT) experienced in everyday life is an illusion because we live in a timeless block universe. On this view, there is a lack of a passage of time when viewed from a physics viewpoint, but a dominance of the passage of time in experienced consciousness, which results in a "two times problem": the relation of *veridical time* (physical time, which does not pass), and *manifest time* (psychologically experienced time, which does) is problematic.

However there is equally a problem for the passage of time in physics by itself. One the one hand there are "block universe" spacetimes such as the maximally extended Schwarzschild, Kerr, TAUB-NUT, and Gödel solutions [3], which are of great interest in terms of demonstrating possible properties of solutions of the Einstein Field Equations (EFE) of General Relativity Theory. However they cannot exist in physical reality because there is no process whereby they can come into being in the context of the expanding universe in which we live; in conceptual terms, they do not fulfil the requirements of Assembly Theory [4]. By contrast physically relevant solutions such as expanding and evolving universe models [5,6], models of the creation of black holes by gravitational accretion processes [7,8], and models of binary black hole accretion accompanied by the emission of gravitational radiation [9,10], represent physical processes taking place over the course of time together with the time evolution of the associated spacetime. The passage of time is central to these processes, with the relevant spacetimes evolving via the EFE. They do not represent block spacetimes.

### 1.1 The main issue

**Block universe spacetimes** are spacetimes that are geodesically complete [3]: they extend as far as is possible spatially and to both the future and the past. Everything that can happen in them has already happened, there is nothing more that can change so they represent an unchanging situation. The claim is that this can as it were be viewed from outside as an unchanging static block spacetime (Fig. 1(a)). An alternative the present author has proposed is an **Evolving Block Universe (EBU)**, which has a future boundary that continually expands to include a larger region of spacetime than before, as time evolves (Fig 1(b)). What is new in this paper is that this proposal is now related, as is the case in [1] and [2], to the way time is experienced by the human mind. This is problematic in the case of a Block Universe in particular because on that view, the future is already determined at every time on a world line: the human mind can make no difference to it. This is not the case in an EBU.

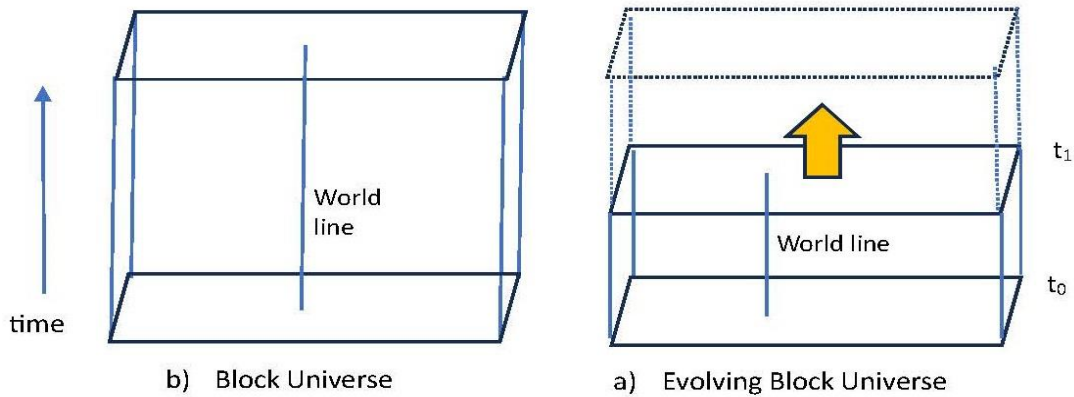

**Figure 1 A Block Universe (Left) and an Evolving Block Universe (Right).** Time is represented vertically. *The Block Universe has no special time: it represents all happenings from the start of time to the end of time, itself being unchanging. The histories of material objects are represented by timelike world lines in it. The Evolving Block Universe at time $t_1$ exists from a starting time $t_0$ only until the present time $t_1$; it is continually extending to the future, as indicated by the yellow arrow. World lines do not yet exist for times later than $t_1$, however they will do so in the future, which is not yet determined.*

The brain can be regarded as an **Information Gathering and Using System** (an IGUS) travelling on a timelike world line, because that is what brains do: gather information, analyze it, and use it to plan future actions, as I discuss in §4. Such systems can exist in either a Block Universe or an Evolving Block Universe. But only in the Evolving Block Universe can they have agency, which they manifestly do [1].

**1.2 Foundations**

    **As regards physics**, the situation is clearly stated by BR in [1]:

*Physics is not the description of static entities: it is the description of processes. The 4-dimensional universe is not an entity, it is a process. Physics is about events, about change. What spacetime describes is a complex network of changes, not a static 4-dimensional block. Upon careful examination, to say that 4-dimensional spacetime is like a "block" is to imagine an additional external time variable, in which the 4-dimensional universe is remaining static. But there is no additional external time variable.*

*The 4-dimensional universe is our map of a multifaceted set of changes. Specifically: to say that the future is "equally real as the present" is an unnecessary redefinition of the word "real", that conflicts with our commonsense use of this word. Relativity does force us to use this word with more caution, but it does not force us to this a-temporal use: we can still say that something is "real here and now", without for this having to say that the "the future is real now".*

Thus in physical reality, things happen and time development takes place. This time evolution is represented by a fundamental set of physics time evolution equations (a) the Dirac equation for the wave function of quantum theory, (b) the 1+3 decomposition of Maxwell's equation determining the time development of the electric and magnetic parts of the electromagnetic field, (c) the similar 1+3 decomposition of Bianchi identities of General Relativity theory determining the time development of the electric and magnetic parts of the Weyl tensor [11,12], and (d) the Arnowitt-Deser-Misner (ADM) equations for tine evolution in General Relativity Theory [13] in general. In all the latter cases there are constraint equations that must be conserved as the time evolution happens.

There are some coordinates which emphasize this time-evolution aspect, and some that hide it.

- *General coordinates* are characterized by general covariance [14,3], and time evolution is hidden when they are used because they are not related to specific physical processes.
- *Symmetry based coordinates* such as those adapted to spherical symmetry (the Schwarzschild vacuum solution and spherical stars) and the spatial homogeneity of the Friedmann-Lemaître-Robertson-Walker (FLRW) solutions will represent time evolution or not, according to the symmetries of the spacetimes (Schwarzschild is static but almost all FLRW models are not).
- *Maximally extended coordinates* represent the geometry of spacetimes in the indefinitely far future when all time evolution has taken place and change has ceased. These are block universes.
- *Harmonic coordinates* are suited to represent radiative processes.
- *Comoving coordinates* [11] are suited to representing the dynamics of relativistic fluids.

**As regards brain function**, BR comment

*For a neuroscientist, time is oriented, always pointing towards the future. We can remember the past (but not the future) and we can influence the future (but not the past). The past no longer exists, the future is open; the present is the only real moment—thus the possibility of time travel to the past is limited to fiction because one cannot travel to a moment that does not exist.*

They end up with the two authors (a physicist and a neuroscientist) essentially agreeing on foundations, but with differences about solutions.

GBM essentially agree with BR as regards the brain. It is because they propose a block universe that the alleged "two-time" problem arises. A solution is proposed by GBM based in the idea [15,16] of an IGUS (Information Gathering and Using System) as the basis of manifest time within the context of a block universe. Such systems arise through natural selection, which begets the "illusory system" for functional purposes.

**1.3 The conundrums**

After the above statement about neuroscience, BR state,

> *For a theoretical physicist, time is more complicated: relativity does not permit an objective notion of a global present, the distinction between past and future requires thermodynamics, hence is statistical only. It is far from obvious why we remember the past but not the future, and why we can influence the future but not the past. There is a sense in which it is easier to think about the whole of spacetime as a single four-dimensional entity (the so-called block universe), in which temporal notions are a matter of perspective. Traveling back in time becomes a subject of theoretical investigation.*

Furthermore, they claim,

**T1)** Relativity is incompatible with an objective notion of a global present.

They also state the well-established fact,

**T2)** All elementary laws of nature that we know are invariant under reversal of the direction of time—a principle referred to as Charge conjugation, Parity inversion, Time reversal symmetry (CPT). These laws include classical mechanics, electrodynamics, quantum theory, general relativity, quantum field theory, and the standard model of particle physics, which all exhibit no arrow of time. However emergent properties such as the Second Law of Thermodynamics and processes of life do exhibit such an arrow. BR associate this time asymmetry only with the thermodynamic arrow of time, which is statistical.

My overall response is given in [17,18,19]: we live in an *Evolving Block Universe* (EBU) which is not a static spacetime maximally extended spacetime extending to future infinity, but rather a spacetime with a time-dependent future boundary that continually extends to the future and thereby establishes a global direction of time. The way this happens is determined by the ADM formulation of General Relativity [13] which makes explicit how spacetime develops as time passes, determining the metric $g_{ab}(x^\mu,t)$ from the matter tensor $T_{ab}(x^\mu,t)$, with the spacetime in turn determining how the matter tensor evolves, once an equation of state is given. The local arrows of time such as those underlying the possibility of biological evolution and brain processes derive from the global Direction of Time in the EBU. The problems T1) and T2) are both resolved by two basic feature: (i) emergence of complexity is always accompanied by symmetry breaking [20]; (ii) specific solutions of physical equations are determined contextually – by boundary conditions, initial conditions, and constraints [21] .

**1.4 This paper**

I discuss in the following, how the Block Universe idea as a proposal for the spacetime in which we actually live is based in a number of misconceptions (§2), with an Evolving Block Universe being a more plausible proposal; the relation between a global direction of time and local arrows of time (§3); multiple interacting timescales in the brain, which is not an ideal clock (§4); the Universe will never be infinitely old (§4); and *Eppur si mueve* (§5).

**2. The Block Universe idea is based on a number of misconceptions**

The block universe model does not describe the real universe (§ 2.1), because of broken symmetries: both Lorentz symmetry and general covariance are broken in the real universe (§ 2.2). Also a maximally extended spacetime does not describe the universe today: it is a manifold with finite past and future boundaries, the latter changing with time (§2.3).

**2.1 A Block universe or an evolving block universe? Contrasting views**

The context for considering the issue is the present day understanding of the expanding universe by working cosmologists [5,6], as summarized in the famous NASA picture (Figure 1). NASA thinks the Universe has an age at present: indeed the Planck collaboration [22] report the age of the universe to be 13:7 Gyr.

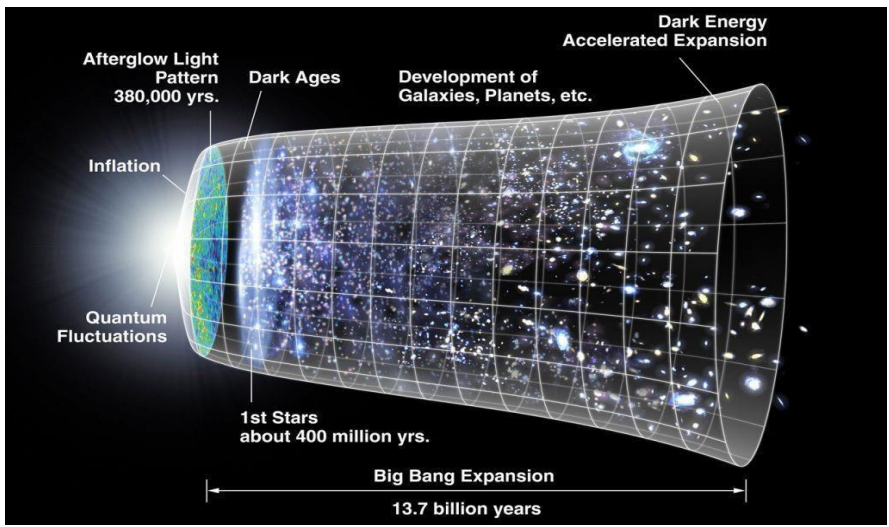

**Figure 2 The expansion of the universe** *In this diagram, time passes from left to right, so at any given time, the universe is represented by a disk-shaped "slice" of the diagram* (Source: Wikimedia Commons). *The present time, corresponding to an age $T_0$ of 13.7 billion years, is represented by the right hand edge.*

The Block Universe viewpoint cannot assign an age to the Universe because then the whole space-time exists as an unchanging single block extending to infinite time, with no preferred present epoch at any time that could be assigned an age (Fig 1(a)). By contrast, in Fig. 2, the present time is defined and represented by the right-hand edge, in agreement with Fig 1(b).

**2.2 Broken symmetries as the basis of emergence of cosmology**

In the real universe, symmetries are broken, as happens in all cases of physical emergence [20]. Underlying the block universe idea are two symmetry related errors.

**A) GR, not SR.** Lorentz invariance and the relativity of simultaneity hold when Special Relativity [14,23] is a good description of spacetime, but do not hold when gravity is significant, as is the case in in cosmology, described by General Relativity and a perturbed FLRW spacetime. The metric tensor $g_{ij}(x^k)$ is determined by the matter present through the Einstein Field Equations [14,3]. It determines proper time τ along any time like world line $x^i(v)$ by the fundamental formula

$$d\tau^2 = - g_{ij}(x^k)\, dx^i\, dx^j .  \tag{1}$$

This time is measured by a clock (an ideal oscillator) travelling on that worldline.

It is an error to assume that the spacetime in which Earth exists is invariant under the Lorentz group, and hence preferred spatial sections cannot exist. This symmetry cannot be used to justify a Block Universe as a representation of the universe in which we live.

**B) Not GR in general, but specific solutions**. At first glance this makes the situation even worse, because general covariance is foundational to general relativity [14,3]: any coordinates whatever can be used so spatial surfaces defining ``constant time'' are then completely arbitrary. But general covariance does not hold usefully in specific cases, where symmetries are broken [24] and specific coordinates are preferred. There are well defined preferred spatial sections in the standard background cosmological models [3], and preferred timelike lines in all plausible cosmological models [5,6].

It is thus an error to assume that because of general covariance, the spacetime in which we live cannot have preferred spatial sections. Physically realistic specific solutions can be represented in any coordinates whatever, but nevertheless generally have preferred spatial sections, and this is the case in cosmology. This symmetry also cannot be used to justify a Block Universe as a representation of the universe in which we live.

**2.3 Not a maximally extended manifold *M*: the present time**

To get a representation of spacetime congruent with the working cosmologist's view as in Figure 1, we need the idea of a manifold **M(t)** with a future boundary **M₊(t)** that changes with time (Figure 2)*.*

The technical concept needed to describe this situation is that of manifold with boundary [25,26]. The real evolving universe is a manifold with a time dependent future boundary. The past exists and the future does not, as in Figure 1 (the spacetime starts at time $t_0$ and comes to an end in the future at time $t_1$ on the left, and at later time $t_2$ on the right in Fig. 2). An earlier manifold $M(t_1)$ is isometrically imbeddable in a unique way as a subset in the later manifold $M(t_2)$, which is therefore an extension of $M(t_1)$. The symmetry group of the FLRW spacetimes determines the physically and geometrically preferred layering via surfaces of constant density, with density changing in time so these surfaces are non-degenerate. These surfaces are also orbits of the $G_6$ isometry group of the spacetime. Worldlines are mapped into each other because they are the timelike integral curves of unique eigenvectors of the Ricci tensor; proper time is determined along them by Eqn. (1). There is thus nothing arbitrary about the map: it is tightly determined by geometry.

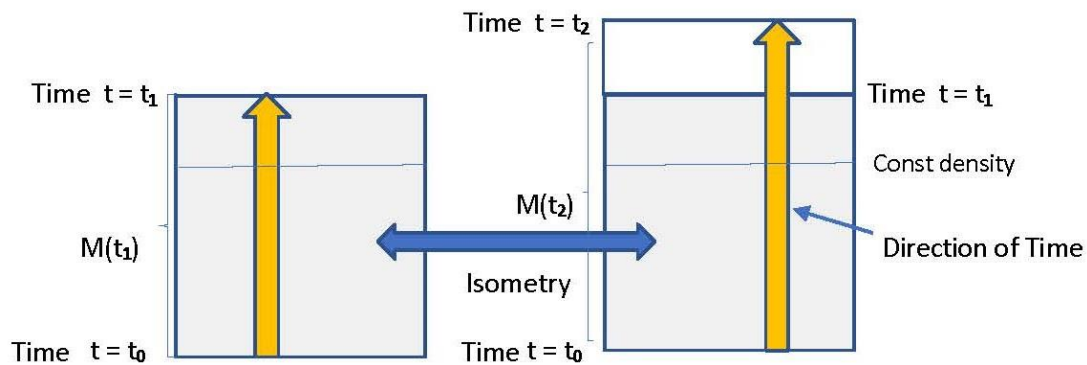

**Figure 3 The Evolving Block Universe and the Direction of Time** *The manifold $M(t_1)$ is the universe at an earlier time $t_1$, with future boundary $M_+(t_1)$ at $t = t_1$, and the manifold $M(t_2)$ the universe at a later time $t_2 > t_1$, with future boundary $M_+(t_2)$ at $t = t_2$. Thus these are the same spacetime depicted at different times. The start of the universe is at the fixed time $t=t_0$. The manifold $M(t_1)$ is isometrically the same as $M(t_2)$ up to the time $t_1$, with uniquely defined surfaces of constant density and fundamental worldlines of matter in the expanding universe mapped into each other by isometries ([3]:43-44): a unique mapping between $M(t_1)$ and $M(t_2)$. The extra domain from $t_1$ to $t_2$ is the amount the universe has grown between those times. The fact that $M(t_1)$ can be isometrically imbedded in $M(t_2)$ uniquely determines that the latter is a later state of the universe than the former. This establishes the global **Direction of Time** pointing from $t_0$ to $t_1$ to $t_2$, whether or not the universe always continues expanding in the future. Recollapse in the future would not affect the Direction of Time.*

Earlier and later states of the evolving universe are uniquely distinguished by the fact that the earlier states are submanifolds of the later states (this is true even if the universe recollapses in the future). This introduces a global **Direction of Time**, pointing from the fixed past boundary (the start of the universe at $t = t_0$) to the continually moving future boundary (the present, at any specific time: in Fig. 2, first at $t = t_1$ on the left and then at $t = t_2$ on the right).

**2.4 Existence of past and future**

**The past exists** because it influences the present, for example nucleosynthesis in first generation stars generated Carbon, Nitrogen and Oxygen that exist on the Earth today. If the past when those events took place did not exist as a physical reality, we would have uncaused entities (these elements of life) on Earth. They have come into being via nucleosynthesis in the interior of stars in our past [5]. These are traces of the past due to entropy gradients (§4(v) in BR). Similarly stars, planets, and human beings exist because of past events. If the past did not exist, we would not be here. We cannot change that past. The underlying philosophical position is that if any entity $\varepsilon$ can be shown to have verifiable physical outcomes in the material universe at the present time, than that entity $\varepsilon$ must also exist, because uncaused entities cannot exist.

**The future does not yet exist** because its outcomes are not yet determined. This is both because (i) of intrinsically random quantum events [27,28], and (ii) the inability to specific initial data to arbitrary accuracy [29-31], leading to unpredictable classical dynamics which can get amplified to macro scales due to the *Real Butterfly Effect* [32]. Chaotic dynamical systems with strange attractors influence the weather [33] so precise outcomes from detailed initial data, which is necessarily subject to the Heisenberg uncertainty principle relating position and momentum, cannot occur. Arguably the same is true for structure formation in the cosmos [34]. Furthermore (iii) agents affect the future by the decisions they make, such as carrying out scientific experiments, or engaging in engineering or artistic projects ([35, 36], §4(v) in BR). The mutability of the future, emphasized by BR, removes one of the arguments put forward for a block universe.

## 3 A global direction of time and local arrows of time

The context of the expanding universe introduces a global direction of time (§ 2.3) which breaks time symmetry [5]. Local basic arrows of time are derived in a downward way in this context (§ 3.1). The laws are time symmetric, but their context of the expanding universe (Figures 1 and 2) is not; that is the source of the broken symmetry leading to local arrows of time. The basic laws then determine derived arrows of time in emergent systems (§ 3.2).

### 3.1 The four basic arrows

An *arrow of time* in physics or biology locally at any time $t_1$ determines a demonstrable difference between the past and the future in terms of physical or biological outcomes at that time, even though the underlying physics is time symmetric. This happens because all physical outcomes are determined in a contextual way [21]: they are not just determined by the underlying fundamental physics, but by the environment within which the relevant system is imbedded. Furthermore, there is more to the local distinction between the past and future than just the thermodynamic arrow of time mentioned by BR. Four basic local arrows of time arise from the Direction of Time in an EBU context [17,18]. They are,

- **The thermodynamic arrow of time, characterized by the Second Law of Thermodynamics.** This results from special cosmological initial conditions in an evolving universe [37-39], together with a dark night sky to act as a heat sink [37,40]. The decrease of temperature with time in the expanding universe hot big bang epoch underlies physical processes largely controlled by thermodynamics: nucleosynthesis, decoupling, and structure formation [5,6], and determines their various arrows of time (helium comes into existence that was not there before, stars and galaxies come into existence that were not there before, and so on).

- **The electrodynamic arrow of time** reflects how electromagnetic radiation is received after a signal is sent, not before, even though Maxwell's equations themselves are time symmetric. Its origin is discussed in [41,42,40]. In an EBU advanced and retarded potential are not equivalent because the future does not yet exist, so there can be no advanced potential and associated Green function as occurs in a block universe. The symmetry between the past and future in Maxwell's equations is thus broken by this cosmological context.

- **The gravitational radiation arrow of time** – gravitational waves are received after they are emitted, not before, despite the EFE being time symmetric - is resolved similarly.

- **Quantum physics is not time symmetric,** even though often being represented as such. In addition to the unitary transformations generated by the Dirac equation, wave function collapse such as occurs in quantum measurements is time asymmetric and so involve a loss of information [43]. Additionally, real quantum measurements [44] are not time symmetric because any physical apparatus interacts with heat baths that affect outcomes [45-47], and thereby get an arrow of time from the expanding universe. This affects the way time underlies processes in physical structures such as digital computers, whose operation at the transistor level depends on these quantum processes [48]. Depressing the key labelled "B" on your computer causes that image to appear on the screen immediately after you press the key, not before.

### 3.2 The derived arrows

These fundamental arrows of time are fundamental to basic physics, but because microphysics leads to the emergent outcomes of macrophysics, biology, and the brain [35], they also lead to emergence of secondary arrows of time in all emergent systems, which are what we experience in daily life. These are derived from the primary arrows of time just discussed. The way emergence of complex systems took place, leading to life, is chronicled in [49]. Layzer comments [40], "*Information is generated whenever the expansion (contraction) rate exceeds the rate of a local equilibrium-maintaining process*". It is through these processes of emergence that the fundamental arrows of time chain up from the basic fundamental processes and lead to derived arrows of time in all emergent systems. Thus there are emergent arrows of time associated with the variety of macroscopic physical and biological processes, all agreeing with the Direction of Time set by cosmology [17].

- **Diffusion**: stirring a liquid leads to mixing, density gradients lead to dissipative fluxes, and so on.
- **Fracture**: dropping an egg or a glass, they break, and cannot be put together again.
- **Waves:** sound waves are heard after they are emitted (I hear you after you speak, not before), tsunamis hit the shore after a sub-sea event, not before, and so on.
- **Chemical processes**: the Second Law of Thermodynamics determines which chemical changes can occur spontaneously as time progresses, from among all possible changes [50].
- **Macromolecular processes** [51] underlie molecular biology in ways embodying an arrow of time, such as reading a stretch of DNA in order to create a specific protein coded by a gene: the protein exists after the gene is read, not before. These processes emerge from the underlying quantum physics in ways explored in [52].
- **Biology: physiological and developmental processes** follow from biochemical ones, underlying the functioning of life, such as breathing, pumping blood, seeing, acting, and so on. The word "function", which is central to life [53], only makes sense in a context where time passes. Above all we age: we do not start with an old body and develop into an infant and then an embryo. We start as infants and end up old and then we die.
- **Biology: Natural selection** that leads to our genetic inheritance begets the functional systems of life, and cannot take place if time does not pass: selected phenotypes and genotypes exist after they are selected [54], not before. This extends to evolution of the conscious brain, capable of experiencing the passage of time, because that capability enhances survival prospects.
- **Brain function** occurs as time passes. Action potential spike chains propagate down dendrites and axons, signaling molecules diffuse across synaptic clefts, neural network strengths are altered via gene regulation, allowing learning to take place, and so on [55], all taking place in one direction of time. This is based in the various secondary arrows of time just discussed.

Lynn *et al* [56] discuss the decomposition of the local arrow of time in interacting systems. There is a cascade down of the basic arrows (§3.1), and a cascade up of the derived arrows [17,18]. Because these derived arrows are all in the end based in the cosmological Direction of Time (Fig. 3), the statement "*For a neuroscientist, time is oriented, always pointing towards the future*" (BR) agrees with what a physical cosmologist determines. One can ask an interesting question here: is there any evidence from astrophysical observations that any of these arrows of time points in the opposite direction somewhere in the Universe? The answer is no.

**4. Multiple interacting timescales in the brain, which is not an ideal clock**

The EBU view I am putting agrees with most of the statements on both BR and GBM about the brain and time. There are a few cases where I disagree. I discuss in this section, Time is basic to brain function: GBM (§4.1); 4.2 Aspects of time: BR (§4.2), Aspects of time: GBM (§4.3)

**4.1 Time is basic to brain function: GBM**

*"The brain is an inherently temporal organ because in many ways its primary function is to learn from the past in order to best predict the future"*. The IGUS view [15,16] of the brain as a system that gathers and uses information is a simplified model of the current predictive processing view of brain function [57-59]. The spacetime view associated with it by Hartle and GBM is an unnecessary add-on: these processes can underlie memory and perception in either a Block Spacetime, or an Evolving Block Universe, and do not of themselves prefer either context.

The hierarchically organized brain [60] is the basis of manifest time via multilevel cortical processing [61,62] with neural networks acting as clocks [63,64]. Timekeeping processes in the brain are due to emergent oscillatory circuits interacting at multiple timescales [65,66}. It takes time for brain processes to happen on various scales, leading to the "specious present". This results in our experience of time: possibly the most fundamental aspect of consciousness and indeed human existence.

**Experienced time** At very short timescales there may be experimentally created postdictive effects (see GBM), but that does not matter for conscious thoughts and decisions and memory effects at the macro scale. The experience of the irreversible passage of time, at timescales of seconds and up, is a fundamental aspect of that consciousness. Insofar as any illusions about this may occur, they are due to the way various brain components interact via oscillatory circuits to generate the sense of consciousness, on the basis of standard physical and chemical reactions that take time to occur. Experiential past, present, and future are not properties of four-dimensional spacetime, but notions describing how individual IGUSs, including humans, process information in any spacetime including an EBU.

**4.2 Aspects of time: BR**

The characterization of different aspects of time in Section 4 of BR is useful. Physical time passes. It is perceived to pass by the brain: perceiving things, adding events to memory, planning what happens and then taking action, and so on.

My only disagreement with this part of their paper regarding the brain is the statement, *"The past no longer exists, the future is open; the present is the only real moment".* The past very much exists in the brain and mind: past events have shaped details of current neural connections [59], which then play a key role in our understandings of events around us and our reactions to them. We are socially situated human beings with past events shaping our current reactions.

### 4.3 Aspects of time: GBM

Their arguments supporting an IGUS interpretation of how we perceive time to pass is fine: it fits in with current neuroscience, and can be accommodated in an EBU context. It does not imply we live in a block universe.

  a) Time is coarse-grained by the brain because of the time brain processes take to complete, and their interactions at various scales. We cannot perceive arbitrary small time intervals.

  b) There is no reason why the brain should be a good clock: the appearance of time passing is an aspect of perception, which is a contex- dependent predictive processing function. GBM give many useful examples. Time may indeed appear to go slow in some circumstances and fast in others, because brain time $T_B$ and proper time $\tau$ are not linearly related [55].

   **Illusions** Where I particularly part company with GBM is the statement "*The phenomenon of dynamism is an experimentally demonstrable illusory experience*". That is an incoherent statement. You can't have any experience whatever if the passage of time is an illusion: there is then no substrate that will make any illusion of any kind possible. Consciousness is real [56] and the experienced passage of time is a key part of the package [57], whether it accurately represents physical time or not. You can have misleading experiences of the passage of physical time due to the contextual nature of perception, as GBM illustrate, but this all takes place in a context where conscious experience is a temporal phenomenon allowed by brain operations taking place as time passes [44]. These are based in local physics and chemistry where time passes with arrows of time inherited from cosmology. Similarly, I do not agree that the "persisting self" is an illusion. The body persists because of physical conservation laws; memories are made physical through brain plasticity [58]. These memories can be retrieved and constitute the core of the mental persistent self: one of the basic features of consciousness.

   It is indisputable that time passes in our experience (you could not be reading this text if that were not so). This is crucial evidence as to the way things work. There is no discordance with physics if we adopt the view put in this paper.

**5: The Evolving Universe will never be infinitely old**

A key feature of the block universe is that in many (most?) cases, they are infinitely old, whereas at any particular time an EBU has a specific finite age. So one might ask, How long will it take for an EBU to become infinitely old? When will it attain the status of a Block Universe?

As stated above, the best estimate of the age of the universe at the present time is $T_o$ = 13:8 Gyr. Suppose our descendants were still alive and did the measurement again after further a time $T_o$ has elapsed, so the Universe then has an age $T_1 = 2\, T_o$. The naïve view is that we would then be closer to the Universe being infinitely old than we were when we first measured the age of the universe at time $t_o$. But, because of the nature of infinity pointed out in [30],

*The Universe will then be no closer to being infinitely old then than it was at time $T_o$.*

(ask yourself, "How much time remains until the end of the universe?" at each of these times). An infinite time will still remain an infinite time to the future at time $T_1$, just as it did at time $T_o$. And the same will be true if we were to measure its age at time $T_N = 10^N\, T_o$, no matter how large $N$. That is the nature of infinity.

Consequently, assuming it expands forever, this will always be true:

*The universe will never be infinitely old.*

(if it does not expand forever but rather recollapses to a future singularity, this will trivially be true).

The fact that we can draw diagrams where infinity is represented as a boundary of spacetime [3] does not mean that the Universe actually ever attains that state. Conformal time misleads us; it does not represent time as measured by any physical clock or system.

6: *Eppur si mueve*

Although the above is a very solid case for an evolving block universe, there will undoubtedly be many physicists who will deny it because they have been deeply immersed in the much-debated "Problem of time in quantum cosmology" [70]. However as already pointed out above, that project is centered on the concept of the existence of a "Wave Function of the Universe": but there are very good reasons to doubt that any such entity exists [52]. Furthermore if it did exist, there is no evidence whatever that it would be governed by the Wheeler-De Witt equation, which is the source of the problem of time in quantum cosmology, and has never been subject to experimental test.

A referee has pointed out four papers [71-74] that make the case for a block universe, which I therefore have to respond to. Paper [71] is a sophisticated defense of the block universe concept based in the relativity of simultaneity. However I have pointed out above that we do not live in a Minkowski spacetime invariant under the Lorentz group. Symmetry breaking is associated with specific solutions of GR, such as the evolving FLRW models which certainly do have preferred surfaces of constant time. Paper [72] is a direct challenge to the position put by BR (see §1.1), denying that physics is to do with dynamical processes. Again the relativity of simultaneity plays a central role in the argument (pp.63-68). Their denial of preferred foliations of spacetimes (p.79) contrasts resoundingly with current cosmological theory [3,5,6] and indeed their own discussion on page 103. They refer to the Einstein equations as an a-dynamical global constraint (e.g. on p.106), which will surely be a surprise to those studying dynamical properties of GR [5-10,13]. Paper [73] is an interesting discussion of the relation between actuality and potentiality, focusing on the question of whether change and temporal passage are real. The author argues that we can know on independent metaphysical grounds—and in particular from the metaphysical presuppositions of any possible physics—that change, temporal passage, and thus the actualization of potential must be real features of the world, which agrees with BR as quoted above. Paper [74] presents a quantum theory of time which is claimed to support a block universe. However it involves *inter alia* the concept of a wave function for a galaxy, but the existence of such a wave function is highly debatable [52]. In any case , quantum physics *per se* does not govern galaxy dynamics [75].

None of these papers provides an argument invalidating the proposals in this paper.

**Abbreviations**

**ADM:** Arnowitt-Deser-Misner [13]
**BR:** Buanomano and Rovelli [1]
**CPT**: charge conjugation, parity inversion, time reversal symmetry
**EBU:** Evolving Block Universe
**FOT**: Flow of time
**EFE**: Einstein Field Equations
**FLRW:** Friedmann-Lemaître-Robertson-Walker
**GBR**: Gruber, Block, and Montemayor [2]
**GR**: General Relativity
**IGUS**: Information Gathering and Using System [15,16]
**SR**: Special Relativity

**Declarations**

The author has no relevant financial or non-financial interests to disclose.
The author received financial support from the University of Cape Town Research Committee

2023/10/05